\documentclass[a4paper,11pt]{article}
\pdfoutput=1
\usepackage{graphicx}
\usepackage{float}
\usepackage{bm}
\usepackage{braket}
\usepackage{amsmath}
\numberwithin{equation}{section}
\usepackage{amssymb}
\usepackage{amscd}
\usepackage{latexsym}
\usepackage{slashed}
\usepackage{color}
\usepackage[normalem]{ulem}
\definecolor{orange}{cmyk}{0,0.5,1,0}
\usepackage{verbatim}
\usepackage{rotating}
\usepackage{cancel}
\usepackage{caption}
\usepackage{hyperref}
\usepackage{cite}
\usepackage[export]{adjustbox}
\usepackage{wrapfig}
\usepackage{mathtools}
\usepackage{makecell}
\usepackage{subfigure}
\usepackage{lineno}
\hypersetup{pdfstartview=FitV,colorlinks=true,linkcolor=blue,citecolor=red,filecolor=black,urlcolor=blue}

\begin{document}

\begin{flushright}

\end{flushright}


\begin{center}

{\large \bf Tagging a Boosted Top quark with a $\tau$ final state} \\    
\vskip 0.6cm

Amit Chakraborty$^{a,}$\footnote{Email: amit.c@srmap.edu.in}, 
Amandip De$^{b,}$\footnote{Email: amandipde@iisc.ac.in},
Rohini M. Godbole$^{b,}$\footnote{Email: rohini@iisc.ac.in},
Monoranjan Guchait$^{c,}$\footnote{Email: guchait@tifr.res.in} 
\vskip 0.3cm

{$^a$Department of Physics, School of Engineering and Sciences, \\ 
SRM University AP, Amaravati 522240, India}
\vskip 0.3cm
{$^b$Centre for High Energy Physics,\\ 
Indian Institute of Science, Bangalore 560012, India}
\vskip 0.3cm
{$^c$Tata Institute of Fundamental Research, \\ 
Homi Bhabha Road, Colaba, Mumbai 400005, India.} 

\end{center}

\vskip 0.3cm

\begin{abstract}
Boosted top quark tagging is one of the challenging, and at the same time exciting, tasks in high energy physics experiments, in particular in the exploration of new physics signals at the LHC. Several techniques have already been developed to tag a boosted top quark in its hadronic decay channel. Recently tagging the same in the semi-leptonic channel has begun to receive a lot of attention. In the current study, we develop a methodology to tag a boosted top quark ($p_T>$ 200 GeV) in its semi-leptonic decay channel with a $\tau$-lepton in the final state. In this analysis, the constituents of the top fatjet are reclustered using jet substructure technique to obtain the subjets, and then $b$- and $\tau$- like subjets are identified by applying standard $b$- and $\tau$-jet identification algorithms. We show that the dominant QCD background can be rejected effectively using several kinematic variables of these subjects, such as energy sharing among the jets, invariant mass, transverse mass, Nsubjettiness etc., leading to high signal tagging efficiencies. We further assess possible improvements in the results by employing multivariate analysis techniques. We find that using this proposed top-tagger, a signal efficiency of $\sim 77\%$ against a background efficiency of $\sim 3\%$ can be achieved. We also extend the proposed top-tagger to the case of polarized top quarks by introducing a few additional observables calculated in the rest frame of the $b-\tau$ system. We comment on how the same methodology will be useful for tagging a boosted heavy BSM particle with a $b$ and $\tau$ in the final state.
\end{abstract}

\newpage
\setcounter{footnote}{0}



\section{Introduction}
Top quark \cite{Kehoe:2007px}, the heaviest fermion in the Standard Model (SM), has a large coupling with the Higgs boson. Hence it plays an important role in many of the suggested ideas of beyond the standard model (BSM) physics, as most of these try to address the issue of radiative stability of the Higgs mass. The top being the heaviest, dominates these radiative corrections. Its coupling with the Higgs boson holds the promise of testing the Higgs sector of the SM and beyond (see e.g.~\cite{BhupalDev:2007ftb,Rindani:2011pk,Rindani:2013mqa,Boudjema:2015nda}) as well as of probing different suggestions/formalisms of BSM physics. Many of these models have a heavy top partner or resonances whose decays involve top quarks and thus provide opportunities to probe these BSM ideas by studying top quark production at the LHC.

Top quarks are of course produced copiously at the hadron colliders, in pairs or singly, respectively, through strong and weak interactions. As mentioned above, another source of $t$-quarks produced at colliders is the decay of hypothetical heavier particles, predicted in various BSM scenarios, to final states containing the top quark. Examples of such BSM models are supersymmetry (SUSY) \cite{Nilles:1983ge, Wess:1992cp,Haber:1984rc,Drees:2004jm, Baer:2006rs} or little Higgs model \cite{Kaplan:1983fs,Schmaltz:2005ky}. Both of these are in fact models which address the hierarchy problem. Models with extra space dimension \cite{PhysRevLett.83.3370} also predict exotic heavy resonances, which would decay to a final state containing one or more top quarks. In all cases the decay vertices are likely to carry an imprint of the BSM in their strength and chiral structure. Hence, a study of the production and decay of the top quarks at colliders provides an excellent avenue to explore BSM physics \cite{Husemann_2017,Beneke:2000hk}. The produced $t$-quark can be successfully tagged in all its decay modes: pure hadronic as well as the semi-leptonic ones, where the $W$ in $t \rightarrow  b W^+$ decays hardonically or leptonically respectively.

In the context of the study of $t$-quarks at colliders, knowledge of the polarization of the produced top quark can provide us with an additional important handle to get information on the interaction vertex. This in turn can shed some light on BSM physics responsible. Luckily, the top decay products can be good polariometers for the decaying quark. This is facilitated by the large mass of the $t$-quark. As a result of this large mass, the $t$-quark, with a lifetime of $\sim 5 \times 10^{-25}$ s, decays before its hadronisation which occurs on a time scale of $1/\lambda_{\text{QCD}} \approx 10^{-24}$~s. As a result, its decay products can carry information about its spin state. In fact, it has been long known that the decay product angular distributions with respect to the spin direction of the decaying top as well as their energy distributions depend on the $t$-polarization \cite{Jezabek:1988ja,Jezabek:1994qs,Bernreuther:2008ju} and hence can be used to gain information about the same. Since $t$-quark polarization is an important probe of BSM physics and the aforementioned correlations follow from the chiral structure of the SM $tbW$ vertex, the effect of anomalous $tbW$ couplings on these correlations also has to be investigated for them to be useful probes of polarization. Such investigations~\cite{Rindani:2000jg,Grzadkowski:1999iq,Hioki:2001eq,Ohkuma:2002iv,Grzadkowski:2001tq,Grzadkowski:2002gt,Hioki:2002vg,Godbole:2002qu,Godbole:2006tq,Godbole:2010kr,Godbole:2018wfy} have shown that the angular distributions of the down type fermion ($\ell, d$), in the decay of the $W$ coming from the top, are particularly robust probes of the $t$-polarization.

With the ever-increasing lower limits on the masses of the BSM particles, the top quarks, expected to be produced in their decays, will necessarily have higher transverse momenta and hence will be highly boosted. This large boost causes the decay products to be highly collimated and these appear in the detector as a single jet in both the hadronic and semi-leptonic decay mode. Tagging these boosted top quarks at the LHC has been an active field of research now for more than a decade.

The opening angle between the decay products depends inversely on the top decay Lorentz factor $\gamma \sim E/m$. The jet which includes all the decay products of the decaying boosted top quark tends to have larger radius than a typical QCD jet which owes its structure to the light parton radiations. The jet corresponding to a boosted top quark is thus a `fatjet' \cite{Salam:2010nqg}. A number of tagging algorithms have been proposed corresponding to hadronic top quark decays using fatjet analysis. These range from those based on substructure \cite{Thaler:2008ju,Kaplan:2008ie,Plehn:2011tg} to the recent ones which include state-of-the-art deep learning methods \cite{Almeida:2015jua,Kasieczka_2019,Kasieczka:2017nvn,Bhattacharya:2020aid,Bhattacherjee:2022gjq}. The jet substructure based taggers make use of the identification of W-boson and top quark through the mass reconstruction while working with jet constituents. The strategies based on jet images have been the main thrust in the deep learning algorithms (for more details, see  Ref. \cite{Larkoski:2017jix,Kogler:2018hem}).

Tagging of a boosted top quark decaying semi-leptonically referred to as leptonic top jet suffers due to the presence of the neutrinos in the final state, which hinders complete reconstruction of the top quark mass. One can of course resort to the transverse mass variables such as $M_{T}$ and  extract the mass of the decaying top quark from the edge of the $M_{T}$ distribution. Alternatively, it is also possible to use the presence of hard tracks originating from the leptons inside a fatjet, to tag\cite{Bhattacharya:2020aid} a leptonically decaying boosted top quark. The authors in \cite{Chatterjee_2020} have shown that it is possible to construct kinematic quantities, which can discriminate between a boosted leptonic top-jet containing non-isolated electrons/muons and a QCD fatjet where light jets are mistagged as leptons.  

In this work, we devise a tagging method to identify boosted top quarks decaying semi-leptonically with a $tau$-lepton in the final state, taking in to account decays of the $\tau$  both in its leptonic and hadronic channels. In addition to the $\nu_\tau$ from the $W$ decay the final state contains one more neutrino coming from the $\tau$ decay as well. This complicates the tagging process. It should be noted that, since the $\tau$-leptons are heavier than other leptons, their couplings and hence polarization are also sensitive to new physics effects. Hence having a top-tagger for a $t$ decaying with a $\tau$ in the final state can open up further possibilities of BSM studies using the polarization of the $\tau$ \cite{Bullock:1992yt,PhysRevD.79.095015,Guchait:2008ar,Gajdosik:2004ed} as well. The major challenge in identifying a $\tau$-jet ($\tau_{h}$), which arises due to the hadronic decay mode of the $\tau$ inside a fatjet, is to distinguish it from quark and gluon-initiated QCD jets. The proposed top tagger relies on efficiently identifying an energetic $b$-jet and a $\tau$-jet within the top quark fatjet. We benchmark the performance of our proposal by using simulated events, corresponding to the production of a heavy $W'$ followed by its decay $W' \rightarrow t b$ and further the decay of the $t$ into a $b \tau \nu_\tau$ final state, applying the jet substructure technique and constructing a few discriminating kinematic observables. We demonstrate that these are very useful in eliminating QCD jets faking as $\tau$-jets and thus facilitate tagging the top jet. We achieve an efficiency of $\sim$ 77\% for tagging the semi-leptonic top quark jet with taus in the final state, while keeping the mistagging efficiencies of backgrounds from light flavour QCD jets to $\sim$ 3\% level.

Even though the main focus of this study is to identify top quark jets with $\tau$-leptons in the final state, the proposed methodology can be applied to any fatjet, which includes a $b$ quark and a $\tau$. For example, a light charged Higgs boson with $m_{H^\pm} < m_{top}$, is an example. In this case the top quark can decay to a bottom quark and a charged Higgs boson which then subsequently dominantly decays through $\tau\nu_{\tau}$ mode. Our proposed methodology can also be used to probe decays of the third generation squark, namely top squark, in R-parity violating (RPV) scenarios.    
In RPV SUSY model, both bi-linear (LH) and tri-linear (LLE, LQD) re-normalizable lepton number violating operators are allowed in the superpotential\cite{Hall:1983id,Drees:2004jm} by gauge invariance and Supersymmetry. In this scenario, RPV decay of top squark $\tilde{t}\to b \tau$, can occur for both the bilinear and trilinear RPV terms. See, for example, \cite{Romao:1999up} and \cite{Chun:2014jha}, respectively. Clearly, our methodology can be used to tag the top squark decaying in this fashion. A third generation Leptoquark of electric charge of +4/3 unit can also have decays to a final state containing a $b$ and a $\tau$ similar to the top quark jet. In these cases, the absence of neutrinos from W decay implies that, unlike the top quark mass, it is possible to reconstruct  mass of Leptoquark or top squark in a straight forward way once we tag the $b$- and $\tau$-subjets efficiently.

This paper is organized as follows. In section \ref{sec:semilep}, we introduce the methodology used for tagging top quarks with taus in the final state and discuss the identification strategy for the $b$- and $\tau$- subjets which are a first step in this study. In section \ref{sec:variables} we introduce several variables which have power to discriminate the signal from the background. The results of the multivariate analysis are presented in section \ref{sec:results}. In section \ref{sec:polarization}, we construct a few polarization sensitive observables to explore the ability of our top-tagger to differentiate between left- and right-polarized top quark. Finally, we summarize in section \ref{sec:summary}.

\section{Methodology : $b$ and $\tau$-jet identification}
\label{sec:semilep}
\begin{figure}[htb!]
	\centering
	\includegraphics[scale=0.2]{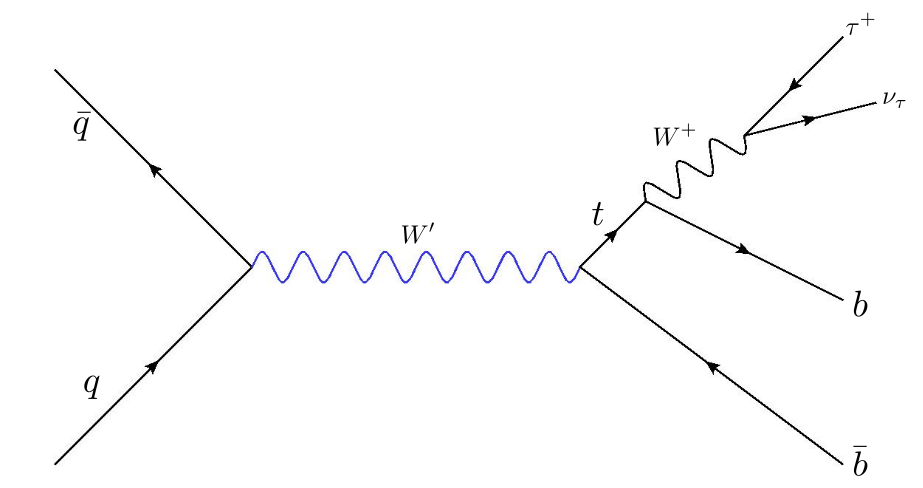}
	\caption{Feynman diagram at leading order for the signal process $pp \to W^{\prime +} \to t \bar{b} \to \tau^+ \nu_\tau b \bar{b}$.}
	\label{fig:feynman_diagram}
\end{figure}

The challenging part of tagging a boosted top quark in the $t \to b \tau^+ \nu_\tau$ decay mode is to identify the $b$- and $\tau$-like subjets inside the top fatjet. In this section, we describe this strategy very systematically. For the purpose of simulation of boosted top quark signal, we generate events for the production of a heavy $W^\prime$ boson with its subsequent decay $W^\prime \to t b$ and further the decay of the $t$ into a $b \tau \nu_\tau$ final state. We do this for the LHC center of mass energy $\sqrt{s}=$13~TeV. Henceforth, we refer to these as $W^\prime$ events. The generated process is indicated in Eq.~\ref{eq:wprime_process}. Even though the Eq.~\ref{eq:wprime_process} and the Feynman diagram correspond only to $W^{\prime +}$ we have of course generated events for $W^{\prime -}$ as well.

\begin{equation}
\begin{split}
pp \longrightarrow W^{\prime^+} \longrightarrow ~&t ~ \bar{b}\\[-0.15 cm]
&\hspace{-0.05cm} \big\downarrow\\[- 0.1 cm]
&\tau^+ \nu_\tau b ,
\label{eq:wprime_process}
\end{split}
\end{equation}
and the corresponding leading order (LO) Feynman diagram is shown in Fig. \ref{fig:feynman_diagram}.

We consider the $W^\prime$ effective model \cite{Sullivan_2002} to generate the signal events setting $m_{W^\prime}$ = 1 TeV. This model is an extension of the Standard Model (SM) incorporating a $W^\prime$ boson with arbitrary vector and axial-vector couplings to the SM quarks. Following \cite{Sullivan_2002}, the relevant part of the Lagrangian can be written as,  
\begin{equation}
\mathcal{L} = \frac{g}{\sqrt{2}} V^{CKM}_{f_i f_j} \bar{f}_i \gamma_\mu (k_R P_R + k_L P_L)  W^\prime f_j + h.c.~,
\label{eq:wprime}
\end{equation}
where $k_R(k_L)$ are the right-handed (left-handed) $W^\prime$ boson gauge couplings to quarks $f_i$ and $f_j$, $V^{CKM}_{f_i f_j}$ are the CKM matrix elements, and $P_{R,L} = (1 \pm \gamma_5)/2$, g is the SM SU$(2)_L$ coupling. For a SM W boson, $k_R=0$, $k_L =1$. Relative values of $k_L$ and $k_R$ determine the polarization of the produced top quark. Its value ${\cal P}_0$ in the rest frame of $\mathrm{W^\prime}$ can be calculated easily by computing $\Gamma^\pm$, the partial decay widths of the $W^\prime \to t_{\lambda} \bar b$ with helicity $\lambda = \pm$ and is given by, 

\begin{equation}\label{eq:pol}
\begin{split}
 \mathcal{P}_0 & = \frac{\Gamma^+ - \Gamma^-}{\Gamma^+ + \Gamma^-} \\
 & = \cfrac{ \left(k_R^2-k_L^2\right) \left(2-x_t^2+x_b^2\right) \sqrt{1+ \left(x_t^2-x_b^2\right)^2 - 2\left(x_t^2+x_b^2\right)}} 
{12 k_R k_L x_t x_b + \left(k_R^2+k_L^2\right)\left(2-\left(x_t^2-x_b^2\right)^2-\left(x_t^2+x_b^2\right)\right)}~,
\end{split}
\end{equation}
with $x_t = m_t/m_{W^\prime},x_b = m_b/m_{W^\prime}$. If we were to neglect the $t,b$ masses, then the produced $t$ will be always left-handed for $k_L=1,k_R=0$ and right-handed for $k_L=0,k_R=1$. If we use Eq.~\ref{eq:pol} to calculate ${\cal P}_0$, for $m_{W^\prime} = 1 $ TeV, $m_t =172$ GeV and $m_b = 4.7$ GeV, and $k_R = 1, k_L =0$ and $k_R = 0, k_L =1$ we get $\pm 0.97 (\sim \pm 1)$,  respectively. Thus the large mass of the $W^\prime$ implies that the polarization of the produced $t$-quark will be decided completely by values of $k_L$ and $k_R$. We will have unpolarized $t$-quarks for $k_L=k_R$. To develop the tagging methodology, we generate unpolarized boosted top quarks from $W^{\prime}$ decay by setting $k_L = k_R $. For the polarization study, we set $k_R$ ($k_L$) to zero to produce left (right) -polarized top quarks.

The $W^\prime$ events ( $pp \to W^{\prime +} \to t \bar b$ ) are generated using {\tt MadGraph aMC@NLO}~\cite{Alwall:2014hca}. It is necessary to take into account the effect of the spin correlations as well as the finite width of the decaying top quark in the decay $t \to b \tau^+ \nu_\tau$. We do this by employing the {\tt MADSPIN} \cite{Artoisenet:2012st} method. Next, $W^\prime$ events are  
passed through {\tt PYTHIA8} \cite{Sjostrand:2014zea} for parton shower and hadronization. In order to consider detector effects, those events are passed through {\tt DELPHES v3.4} \cite{deFavereau:2013fsa} with Compact Muon Solenoid (CMS) card setting. The process of identifying boosted top jets begins with construction of fatjets, setting jet radius parameter $R$ = 1.5 and anti-$k_T$ jet clustering algorithm \cite{Cacciari_2008} as implemented in {\tt FastJet v3.2.1} \cite{Cacciari_2012}. Then, we select those fatjets which pass the threshold of transverse momentum, $\mathrm{p_{T_J}^{min} = 200 ~GeV}$. The boosted fatjets are contaminated by several sources, such as soft radiation, underlying events and multi-particle interactions. These are removed by applying {\tt SoftDrop} technique \cite{Larkoski_2014} setting free parameters $\beta$ and $z_{cut}$ to their values for the standard CMS choice \cite{CMS:2020poo}, viz. $\beta=0$ and $z_{cut}=0.1$. The constituents of the soft-dropped fatjet are further re-clustered with jet radius parameter $R$ = 0.5 using anti-$k_{T}$ algorithm to form subjets with a minimum $\mathrm{p_{T_j}^{min} = 20 ~GeV}$. In principle, $W^\prime$ is produced at almost rest; hence $W^\prime$ decay to $b$ and $t$ will produce two back-to-back fatjets. For the above mentioned choices of various parameters, we find that for $m_{W^\prime}=1$ TeV about $ 58\%$ events contain two back-to-back fatjets whereas the fraction of these events rises to $\sim 79\%$ for $m_{W^\prime}=3$ TeV. To ensure that a fatjet is indeed a topjet, it is necessary to identify the subjets in it as $b$- and $\tau$-jets. We will describe our strategy for identification of the $b$- and $\tau$- like subjets after describing the different event samples we generate for the backgrounds as well.

High $p_T$ QCD jets can fake as top jets. Hence, we need to study the impact of jets produced via QCD processes while developing the strategy to tag top jets in the semi-leptonic channel. The kinematics of the top decay looks similar to QCD parton splitting when its boost factor $y_{t}$ is $\sim ~ 1/\alpha_{s}$. Therefore, at high $p_T$, the subjets of the QCD fatjet, are very likely will be misidentified as $b$- or $\tau$- jets. Hence, efficient identification of the subjets as $b$- and $\tau$- jet is necessary to suppress the number of QCD events significantly. Various properties of the QCD jets differ significantly from the top-fatjet and the idea is to exploit these differences to reduce the QCD contamination. One such property is the average mass of the jet, which is affected mainly by the sharing of energy among different members of the jet. For a QCD fatjet this average mass increases with the $p_T$ of jets. It has been shown that for QCD jets with $p_T$ more than $\sim$ 300 GeV, the corresponding mean jet invariant masses lie within the window of 30-160 GeV \cite{Thaler:2008ju}. This mass range, covering the semi-leptonic top mass, is precisely the one that is important for our  analysis which aims to tag a top quark.

QCD jets have a steeply falling $p_T$ distribution, but due to the much larger production cross-section of the jets, there is a significant number of jets even after the requirement of large minimum $\mathrm p_T$ of $200$ GeV for the jet. The steeply falling $p_T$ spectrum of these jets means that one has to take extra care to generate appropriately large number of events. QCD events from $pp \to jj$ are generated in three $p_T^{\text{bins}}$: [200-300, 300-600, $>$ 600] GeV where $p_T$ corresponds to the transverse momentum of hard scattered particles in the final state. The number of simulated events is determined by keeping in mind that we perform our analysis for a luminosity of 10 $\text{fb}^{-1}$ and we need to pay particular attention to the region [300-600] GeV, since the QCD jets in this $p_T^{\text{bin}}$ contribute dominantly to a window in jet mass which is populated by the signal from semi-leptonic decay of the top quark. We have simulated 10M, 7M and 3M events in these three $p_T^{\text{bins}}$, respectively. We also need to simulate the hadronically decaying top-antitop quark pair ($t_h \bar t_h$) events which is a potential background. The list of generated events used for simulation is presented in Table~\ref{tab:sample} along with the range of $p_T^{\text{bins}}$ for the dijet QCD events as well as for the $t_h \bar t_h$ events. Next, we proceed to discuss the identification of the subjets as $\tau$- and $b$- subjets, which, as mentioned above, is very important so as to be able to handle the background from QCD jets. 

\begin{table}[h!]
	\centering
	\begin{tabular}{ |c|c|c| } 
		\hline
		Process & $p_T^{\text{bins}}$ (GeV) & Cross Section (pb) \\ 
		\hline
		& 200--300 & $\approx 4.9 \times 10^{4}$ \\ 
		Dijet QCD & 300--600 & $\approx 7.9 \times 10^{3}$ \\ 
		& $>$ 600 &  $\approx 2.0 \times 10^{2}$ \\ 
		\hline
t$\overline{\text{t}}$ Hadronic($t_h \bar{t}_h$) & $>$ 150 GeV & $\approx$ 208.2 \\ 
		\hline
		$\text{W}^\prime \rightarrow t\bar{b}\rightarrow \tau b \overline{b}\nu$ & Mass of $\text{W}^\prime$ = 1 TeV &  $\approx 0.6$ \\
		\hline
	\end{tabular}
	\caption{List of the signal and background samples used with their kinematics.}
	\label{tab:sample}
\end{table}

\vspace{0.5cm}

\noindent $\bullet$ {\large $\tau$-jet identification:}
\vspace{0.5cm}

The $\tau$-lepton decays hadronically with a probability of ~65\% producing charged (mainly $\pi^\pm$) and neutral hadrons ($\pi^0$). Hence the multiplicity of decay products, particularly charged tracks, is low in numbers and they are highly collimated in a cone with $\Delta R = \sqrt{\Delta \eta^{2} + \Delta \phi^{2}} < 1.5 $, where $\Delta \eta$ and $\Delta \phi$ are the differences of pseudo rapidities and azimuthal angles respectively between two particles. In collinear approximation, i.e. if $p^{\tau}_{T}$ $>>$ $m_\tau$, the decay products are collimated even more so as to be contained in a jet of even smaller radius, say $R<~0.5$. Among the $\tau$ decay products, the neutral pions deposit a considerable fraction of electromagnetic energy in the calorimeters through photon. This is accompanied by one or three-prong low $\mathrm{p_T}$ charged track multiplicity observed in the tracker, which are the characteristics of a $\tau$-jet. Currently, ATLAS and CMS have developed very sophisticatedly dedicated algorithms for the identification of $\tau$-jets using attributes of $\tau$ decay, such as energy difference in calorimeter cells, lifetime and mass, track multiplicities etc. \cite{Bagliesi:2007qx,Katz:2010iq,Gennai:2006irx}. However, those techniques are beyond the scope of our present analysis. 
Instead, we use a naive track-based isolation algorithm to identify $\tau$-like subjects.

\begin{figure}[htb!]
    \centering
	\includegraphics[scale=0.34]{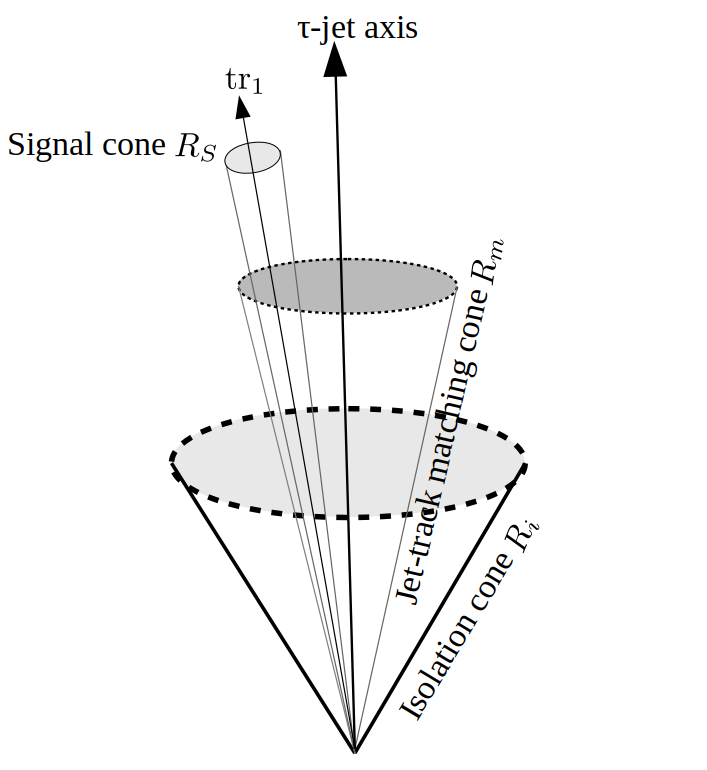}
	\caption{Basic principle of $\tau$-jet identification with charged track isolation.} 
    \label{fig:1}
\end{figure}

In this procedure \cite{Bagliesi:2007qx}; first, we identify the tracks which are within the jet-track matching cone with radius $R_m = 0.1$ calculated using the candidate $\tau$-jet axis, and then select those tracks with minimum $p_T$ of 2 GeV (see Fig. \ref{fig:1}). Among these tracks, we identify the leading track (or the seed track) with minimum $p_T>$~6 GeV and $|\eta|$ $<$ 2.5. Other tracks are accumulated in a narrow region around the seed track with a cone radius $R_{s} = 0.07$, and we demand that the difference of the z-impact parameter $\Delta z_{tr}$ between the leading track and these selected tracks (signal cone) is smaller than 2 mm. This additional requirement ensures that all the tracks in the signal cone are coming from the same $\tau$-lepton decay. 
\begin{figure}[htb!]
	\begin{minipage}{0.48\textwidth}
		\centering
		\includegraphics[scale=0.27]{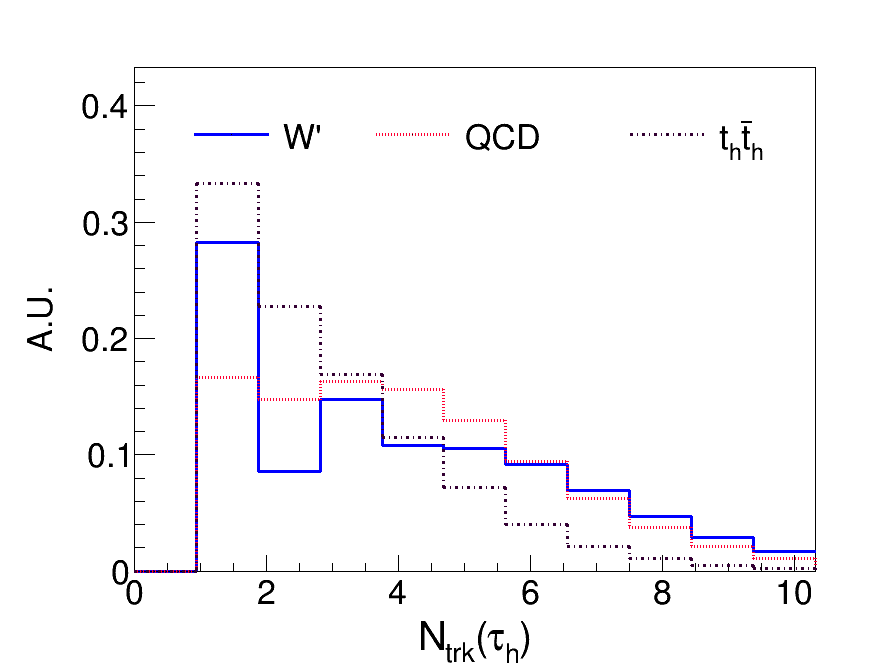}
	\end{minipage}\hfill
	\begin{minipage}{0.48\textwidth}
		\centering  
		\includegraphics[scale=0.27]{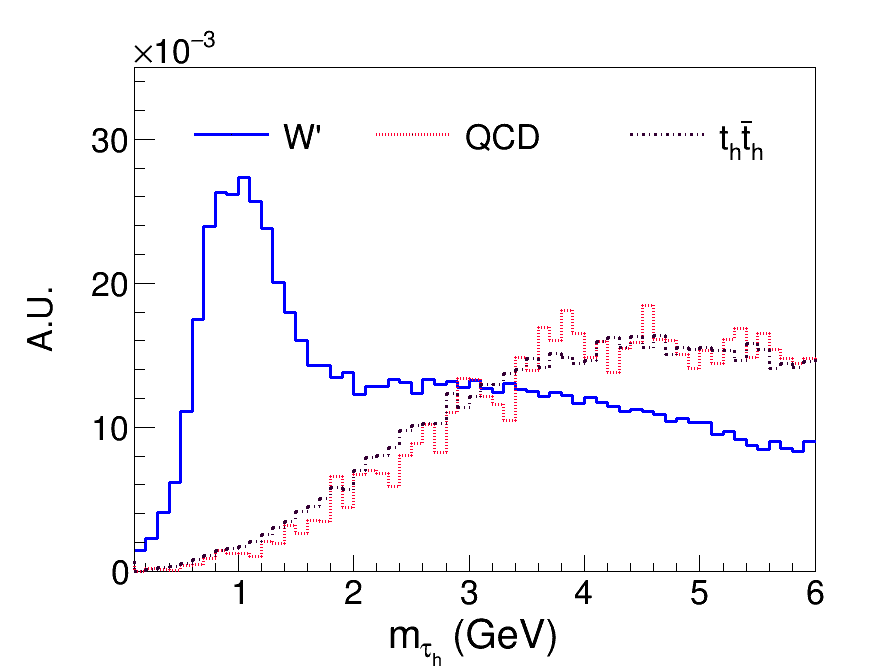}
	\end{minipage}
	\caption{Distribution of the number of charged tracks inside the $\tau$-subjet (left) and mass of the $\tau$-subjet (right) for signal ($W^\prime$) and background events. The solid blue line shows the distribution for the signal while the same for QCD and $t_h \bar{t}_h$ are shown in red dotted and black dash dotted lines, respectively.}
	\label{fig:tau_mass_prong} 
\end{figure}
Furthermore, we adopt a track isolation method, where isolated tracks are reconstructed in a larger cone size $R_{i} =0.45$ around the candidate $\tau$ jet axis with minimum transverse momentum $p_T^i$ of 1 GeV. The isolation criterion is satisfied when the number of tracks (1 or 3) in the isolation region is the same as in the signal cone. Following this naive technique, we achieve $\tau$-identification efficiency $\epsilon_{\tau} \sim ~ 60\%$ for a moderate range of $p_T=20-60$ GeV of $\tau$-jets. In case of QCD, the misidentification efficiency of $\tau$-jets is $\sim$ 5-6$\%$.   

The effectiveness of the $\tau$ identification method is demonstrated in  Fig.~\ref{fig:tau_mass_prong} in distributions of multiplicity of the charged tracks $\left( \mathrm{N_{trk}}(\tau_h)\right)$, mass ($m_{\tau_h}$) for the identified $\tau$-jets. The left plot of charge track multiplicity shows a clear tendency to one and three prong structures for identified $\tau$-jets, as expected. QCD jets acquire mass through multiple splitting and the EM clusters are far away from the jet axis than that of signal, therefore, a smeared distribution is more likely.
While a clear peak structure is visible in the mass distribution for the signal events, shown by the solid blue line (right plot in Fig. \ref{fig:tau_mass_prong}), the invariant mass distribution for the misidentified $\tau$-jet shows a long tail, for the QCD jets as well as the jets from $t_h \bar t_h $, displayed in the same plot by red dotted and black dash dotted lines respectively. It is true that the invariant mass distribution for the signal events does have a somewhat long tail. However, it is possible to reduce it further by using isolation criteria in addition to the single one that we have used \cite{refId0}.

\begin{table}[bht!]
        \centering           
		\begin{tabular}{ |c|c| }
		\hline
		$\mathrm{\mathrm{\mathrm{p_T}}}$ of the $b$-jet (GeV) & $b$ tagging Efficiency \\ 
		\hline
		upto 60 & 60\% \\ 
		\hline
		[60,200] & 80\% \\ 
		\hline
		[200,400] & 60\%\\
		\hline
		 400 - & 50\%\\
		\hline 
	\end{tabular}
	\caption{$b$-tagging efficiencies for different $p_T$ range of the jets following the performance of ATLAS $b$-identification algorithms with Run2 data \cite{Aad:2019aic}.}
    \label{tab:eff_b}
 \end{table} 
 
\vspace{0.7 cm}
\noindent $\bullet$ {\large $b$-jet identification:}
\vspace{0.5cm}

A subjet inside the candidate top fatjet is identified as a $b$-jet, 
if the angular distance $\Delta R$ between the jet and the nearest B-hadron satisfy $\Delta R < 0.5$. For signal events, B-hadrons dominantly arise from the $b$-quarks, which are produced through the decay of $W^\prime$ and the $t$. Intuitively, the B hadrons, which come from the $b$ quark originating from the top quark decay, are likely to satisfy the above matching condition viz. $\Delta R < 0.5$. On the other hand, for QCD multijet events, B hadrons mostly originate due to a gluon splitting into a $b \bar b$ pair and hence unlikely to be close to the fatjet axis.
In $b$-jet identification strategy, we also take into account the impact of detector effects by incorporating the $b$-tagging efficiencies and mistag rates reported in \cite{Aad:2019aic} and which are summarized in Table \ref{tab:eff_b}. Following the CMS analysis, we have used a mistag rate to be $2\%$ for a light jet to be identified as a $b$-jet, 
irrespective of $p_T$ of the jet~\cite{CMS:2012feb}. In our simulation, we correct the $b$ identification probability by applying all these efficiencies. We get $\sim 77\%$ identification efficiency for $b$.

\begin{figure}[bht!]
	\centering
	\includegraphics[scale=0.27]{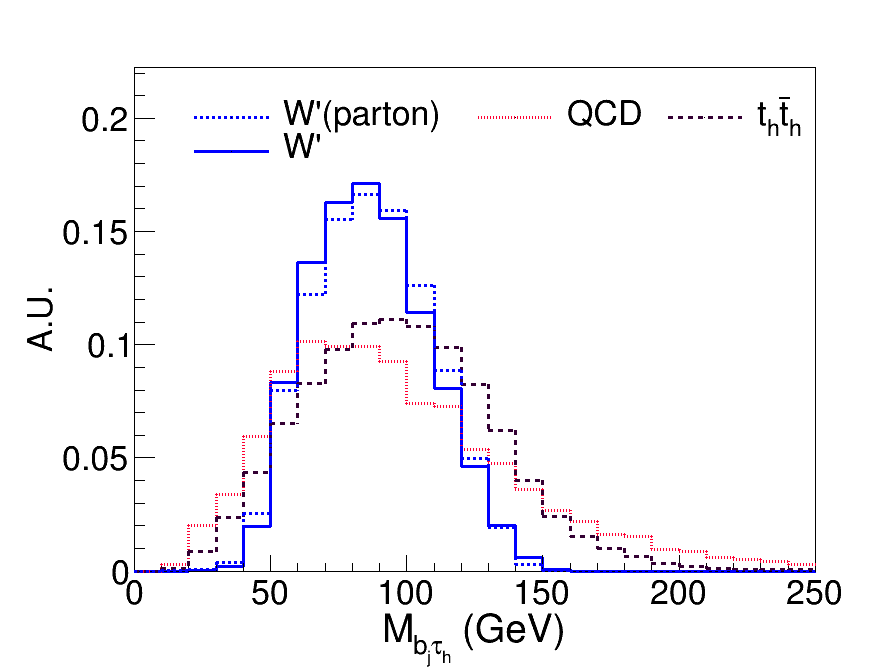}
	\caption{Distribution of invariant mass of $b-\tau$ jet system for signal and background events. The parton level (blue dotted line) curve for $W^\prime$ is constructed out of momenta of the b-quark and visible decay products of $\tau$. The color code for all the other curves is the same as in Fig.~\ref{fig:tau_mass_prong}.}
	\label{fig:2}
\end{figure}

Fig.~\ref{fig:2} shows distributions in the invariant mass of the fatjets containing $b$- and $\tau$-like subjets, identified to be so, using the above procedure for all the three types of events (Table~\ref{tab:sample}) that have been generated, viz. $W^\prime$ and hadronic top and QCD events.
For the sake of comparison, the same invariant mass constructed out of the total four-momentum of $b$-quark and the visible decay products of the $\tau$-lepton ($\tau^{vis}$), referred to as $b-\tau^{vis}$ system is shown. At the truth level, mass of $b-\tau^{vis}$ system $m_{b\tau^{vis}}$ is expected to be bounded by $m_{b \tau^{vis}}  < \sqrt{m_t^2 - m_W^2} \approx 152$ GeV in the limit $m_{b} \to 0 $. A clear peak is visible in the distribution for parton level events around $\sim$ 75~GeV. The reconstructed top jet mass distribution from $W^\prime$ event is found to have a peak too around the same value and the distribution is smeared due to the hadronization of the $b$-quark and consequent decay of the $B$-meson as well as the decay of the $\tau$ lepton and the detector effects. Notice that the distribution does not show any peak-like structure for the $t_h \bar t_h$ and QCD events. This indicates to us that a window in the distribution of invariant mass of the $b$- and $\tau$-subjets, i.e. $M_{b_j\tau_h}$, say between $60$ to $160$ GeV, could be chosen optimally to tag a semi-leptonic $t$-fatjet.

\section{Top jet identification}
\label{sec:variables}
In this section, we construct a number of observables that can be used to identify top jets. We do this by exploiting the features of the $b$- and $\tau$-like subjets that we observed in the earlier section. 
\begin{itemize}
\item{\bf \underline{Transverse Mass}}

The dominant source of missing transverse energy (MET) in the signal is due to the presence of two neutrinos: one from decay of $W$ and the other from the decay of the $\tau$ as can be seen from Eq.~\ref{eq:wprime_process}. The presence of two neutrinos makes it difficult to reconstruct the mass of the top fatjet, compared to the case of the hadronically decaying $t$, where such reconstruction plays an important role in its tagging.
However, the direction of MET can help to do the job in the present case because of the boosted nature of the $t$ and the $W$.  In fact, as a result of the boosted nature of the $t$, $W$ and $\tau$, the direction of MET is expected to be collinear to $\tau$ decay products. This feature can be utilized to get an additional handle to identify top jets and also to reduce background. Keeping the above kinematics in mind, we require the $\Delta R$ between MET and the candidate top jet to be $< 3$. We construct a transverse mass observable, combining the momenta of the identified $b$- and $\tau$-jet, along with the MET given by,  
\begin{equation}
 m_T^2 = m_{b\tau^{vis}}^2 + 2(E_T^{b\tau^{vis}} {\cancel E}_{T}^\nu - 
\mathbf{p}_T^{b\tau^{vis}} . {\mathbf{\cancel p}}_{T} ),
\label{eq:1}
\end{equation}
where $m_{b\tau^{vis}}$ and $\mathbf{p}^{b\tau^{vis}}_{T}$ denote the invariant mass and transverse momentum of the $b-\tau^{vis}$ system respectively. The transverse mass $m_T$ of $b$-$\tau$ jet, displayed by the solid blue curve in Fig. \ref{fig:transverse_mass} is expected to have an endpoint at $t$ mass which is smeared due to difference in correlation of MET coming from two neutrinos with top jet as well as other effects mentioned in the previous section. We can use this feature of the $m_T$ distribution to select the candidate fatjet after requiring an invariant mass of $M_{b_j\tau_h}$ in the aforementioned mass window of [60-160] GeV, which removes a good fraction of background events.

\begin{figure}[tbh!]
	\centering
	\includegraphics[scale=0.27]{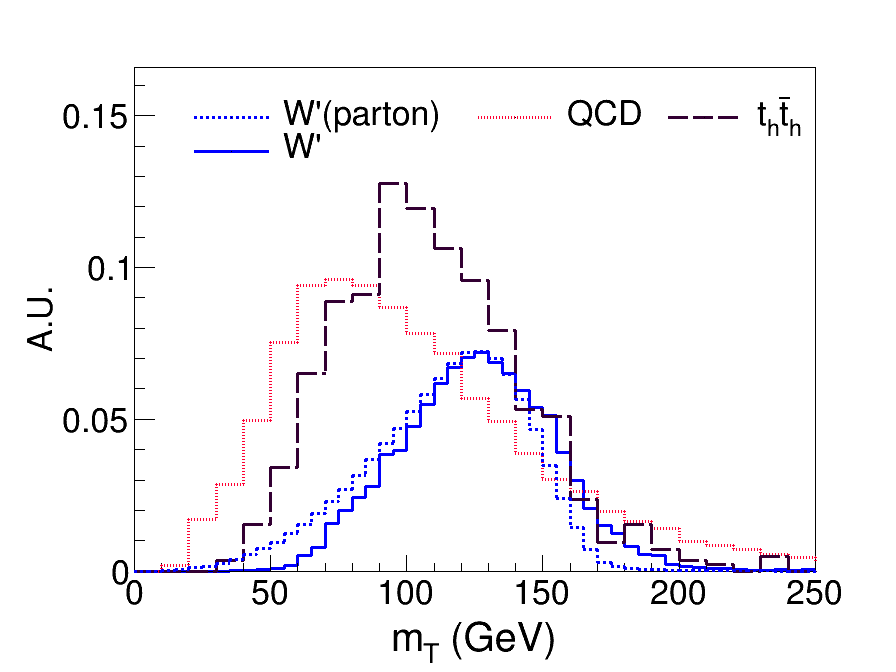}		
	\caption{Distribution of the transverse mass constructed out of $b$-subjet ($b_j$), $\tau$-subjet ($\tau_h$) and MET for the signal and background events. The color code is the same as in Fig.~\ref{fig:2}.}
	\label{fig:transverse_mass}
\end{figure}

\item{\bf \underline{Energy sharing of subjets}}

The pattern of energy sharing of subjets is very different for the signal and backgrounds. This facilitates the construction of an observable which offers good separation between boosted top jets from the signal and the background. The fraction of energy carried by a subjet (j) of a fatjet (J) as  
\begin{equation}
    Z_j = \frac{E_j}{E_J}. 
\label{eq:efrac}
\end{equation}

\begin{figure}[htb!]
	\begin{minipage}{0.48\textwidth}
		\centering
		\includegraphics[scale=0.27]{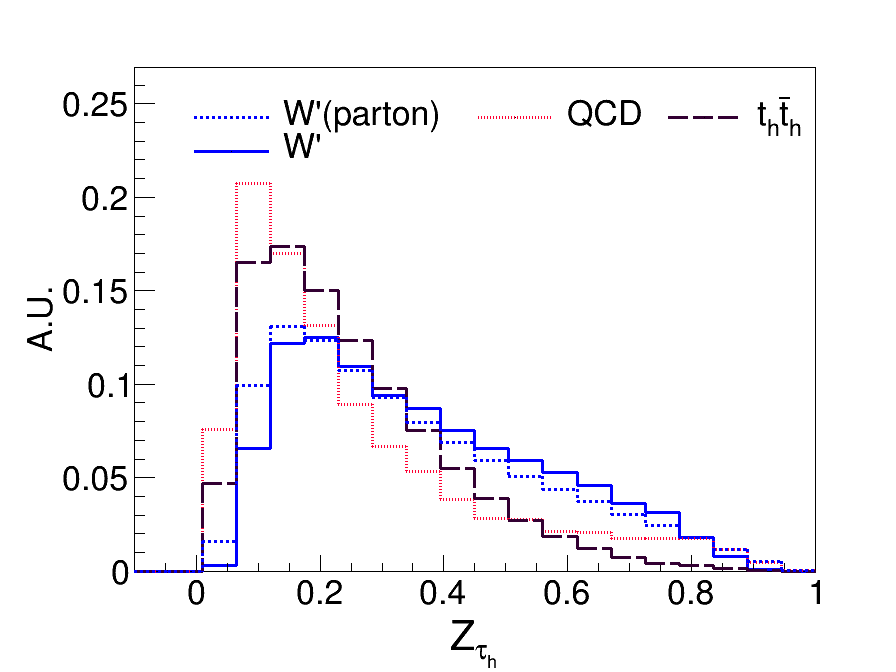}
	\end{minipage}\hfill
	\begin{minipage}{0.48\textwidth}
		\centering  
		\includegraphics[scale=0.27]{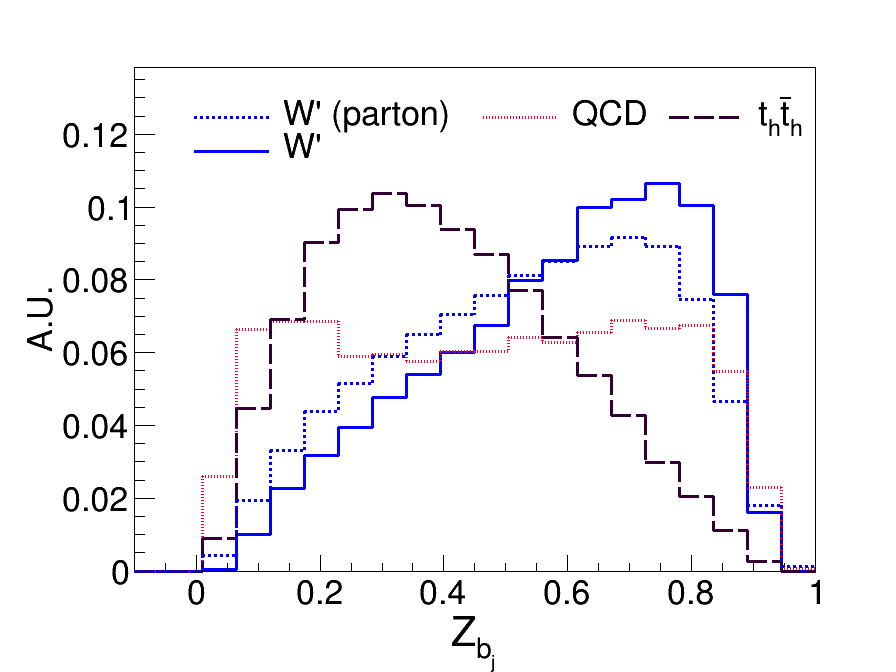}
	\end{minipage}
	\caption{Distribution of energy fraction of $\tau_h$ (left) and $b_j$ (right) in top jet for signal and background events. The color code is same as Fig.~\ref{fig:2}.}
	\label{fig:efrac}
\end{figure}

In Fig.~\ref{fig:efrac}, we show the $Z_j$ distribution for $\tau_h$(left panel) and $b$-like(right panel) subjets corresponding to both signal and background events. For one-to-one correspondence with the jet level, the energy fractions at the parton level are defined as the ratio of the energy of the $b$ quark or energy of the visible decay products of the $\tau$ to the sum of the two ($E_b + E_{\tau^{vis}}$).
For signal events, it is expected that the $b$-subjet will carry a large fraction of energy while $\tau_h$ will share comparatively a smaller fraction of energy of the top system since a fraction is taken away by MET which contains the neutrino from the $\tau$ decay. One can see that $Z_{\tau_h}$ peaks around 0.2. 
However, for background events, these energy fractions of the $b$- and $\tau$- like jets are uncorrelated. In fact, $Z_{b_j}$ in QCD, the dotted red line in the right plot of Fig.~\ref{fig:efrac}, shows a flat distribution, whereas $Z_{\tau_h}$ for a QCD jet mimicking a $\tau-$ like subjet is dominantly distributed to much lower values. The peculiar sharing of energies, particularly for $Z_{b_j}$ in QCD events, can be attributed to additional components the subjet contains due to soft radiation, which could not be removed even after the application of soft drop method \cite{Larkoski_2014}. 

We exploit this characteristic of background sub jets by defining a new variable, $\Delta X_{b_j\tau_h}$ which includes the ratio of the mass of the $b$- and $\tau$- subjet system ($M_{b_j\tau_h}$) with respect to the mass of the corresponding identified candidate top jet ($M_J$),
\begin{eqnarray}
\Delta X_{b_j\tau_h} = 1 - \frac{M_{b_j\tau_h}}{M_J}. 
\label{eq:mass_frac}
\end{eqnarray}
Clearly, for signal events, it is expected to peak around $\Delta X_{b_j\tau_h} \sim0$ as shown in the blue line of Fig.~\ref{fig:mass_frac}. For the backgrounds, as explained above, $b$-jets are contaminated by soft components and hence are much flatter and extend to larger values. We find that this variable helps improve the top tagging efficiency. 

\begin{figure}[htb!]
	\centering
	\includegraphics[scale=0.27]{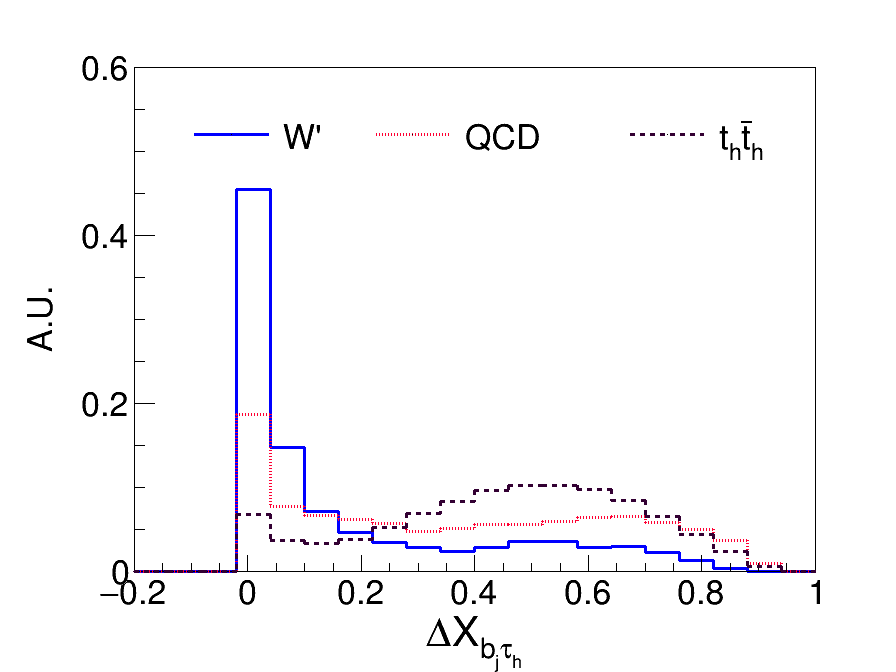}
	\caption{Distribution of variable $\Delta X_{b_j\tau_h}$ (Eq.~\ref{eq:mass_frac}) for signal and background events. Colour code is the same as in Fig.~\ref{fig:tau_mass_prong}.}
	\label{fig:mass_frac}
\end{figure}
 
\item{\bf \underline{N-subjettiness}}

Another very useful variable in the jet substructure technique, which helps to improve our study is N-subjettiness. Subjettiness \cite{Thaler_2011} takes advantage of the different energy flow in the different particles present within the fatjet. It effectively counts the number of subjets in a given jet. 
If there are N candidate subjets in a specific jet, one calculates subjettiness as,    
\begin{equation}	 
\tau_{N} =\frac{1}{\sum_{k} p_{T,k}R_{Jet}} 
\sum_{k} p_{T,k} \,\text{min} \{ \Delta R_{1,k}, \Delta R_{2,k}, ...., 
\Delta R_{N,k} \}.
\label{eq:n_subjetti}
\end{equation}
Here k runs over all the constituents of a jet of momentum $p_{T,k}$ and $\Delta R_{J,k} = \sqrt{(\Delta \eta)^{2} + (\Delta \phi)^{2} }$ is the distance in $\eta - \phi$ plane between a candidate subjet j and a constituent k. The normalization factor is taken to be the jet $p_T$ multiplied by its radius. In the limit $\tau_N \rightarrow 0$, the jet must 
have $\Delta R_{N,k} =0$ i.e all the radiation is perfectly aligned along the candidate subjets, and therefore the jet has exactly N subjets. In case of $\tau_N \rightarrow 1$, the jet must have a large fraction of its energy distributed away from the candidate subjet direction, therefore it has at least N + 1 subjets i.e the minimization missed some subjet axes.


\begin{figure}[htb!]
	\begin{minipage}{0.48\textwidth}
		\centering
		\includegraphics[scale=0.27]{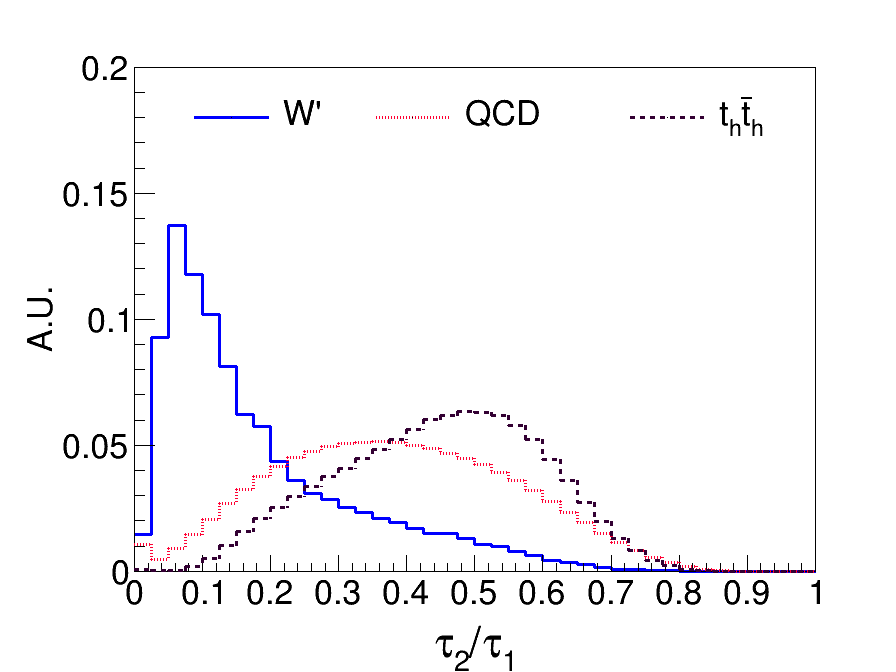}
	\end{minipage}\hfill
	\begin{minipage}{0.48\textwidth}
		\centering  
		\includegraphics[scale=0.27]{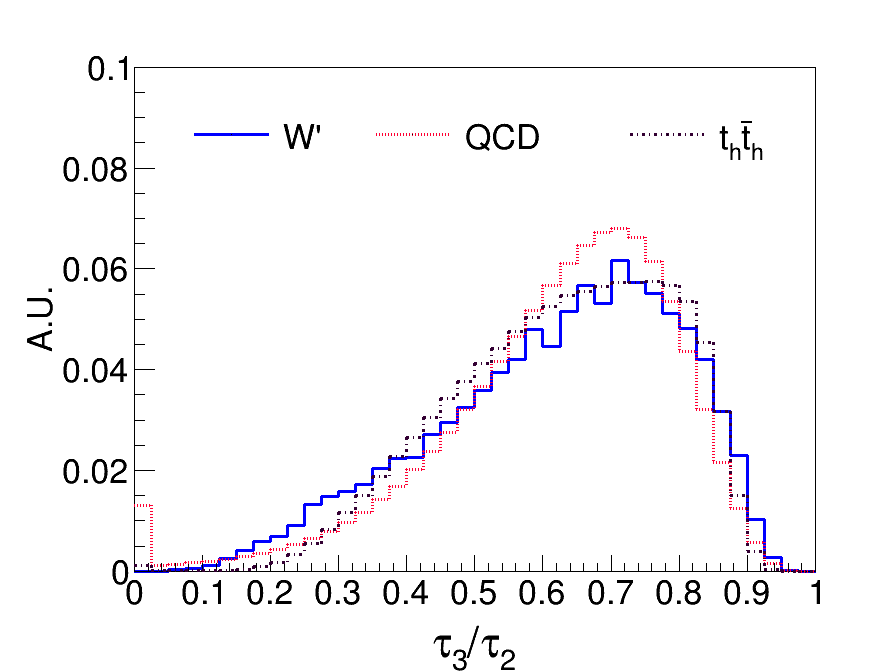}
	\end{minipage}
	\caption{Distribution of $\tau_{2}/\tau_{1}$ (left) and $\tau_{3}/\tau_{2}$ (right) for leptonic top and background jets. Discriminators like $\tau_{2}/\tau_{1}$ measure the relative alignment of the jet energy along the individual subjet directions. Colour code is the same as Fig.~\ref{fig:tau_mass_prong}.}
	\label{fig:jetty}
\end{figure}

Therefore, jets with smaller $\tau_N$ are said to be N-subjetty, whereas larger $\tau_N$ have more than N subjets. QCD jets can have larger values of subjetiness variables due to diffuse spray of large angle radiation, and hence individually $\tau_{1}$ or $\tau_{2}$ can not provide much distinction between signal and background events. The ratio $\tau_{2}/\tau_{1}$, however, is a different story and can be effective in discriminating the different two-prong objects. Similarly, $\tau_{3}/\tau_{2}$ is a better choice while probing three-prong objects. We show the distribution of these two variables viz., $\tau_{2}/\tau_{1}$ and $\tau_{3}/\tau_{2}$ in the left and right panel of Fig.~\ref{fig:jetty} respectively. For the leptonically decaying $t$-quark the fatjet is essentially a two prong object and hence $\tau_{2}/\tau_{1}$ is found to be a good discriminating observable.

\end{itemize}

\noindent Based on the various observations in the discussion above we list the steps to be followed systematically to identify a boosted top  jet in its decay channel, $t\to \bar b \tau^+ \nu_\tau$. 
  \begin{enumerate}
  	\item Cluster the final state hadrons of the events to jets with a minimum $p_T$, say $p_{T} \sim$ 200 GeV or larger, setting the jet size parameter R = 1.5. Then apply a jet grooming technique (e.g., SoftDrop) and remove the soft and wide angle radiation contamination. Select the fatjet mass with a window of the range 60-160 GeV. It is likely to be a top candidate jet.  
  	
  	\item Recluster the constituents of the candidate fat top jet to a smaller radius ($ R \sim 0.5$ ), and select those events where the candidate jet consists of at least two subjets.  
  	
  	\item Identify $b$- and $\tau$- like subjets following the procedure as described in Sec 2. 
    
    \item 
    Finally, if at least two different subjets in close proximity originating from the candidate fatjet pass the $b$ and $\tau$ jet identification, the fatjet can be considered as a top-fatjet if the invariant mass lies in the window 60-160 GeV.   
    
 \item 
    Furthermore, construct the following discriminating observables to reduce possible backgrounds. \\
    (a) Transverse mass$(m_T)$, constructed from the $b$-$\tau$ subjet system and MET following  Eq.~\ref{eq:1}. \\
    (b) Energy Fractions of the $b$- and $\tau$- subjets inside the top fatjet. One can then utilize the difference in the template of energy sharing between the subjets ($Z_{b_j}, Z_{\tau_h}$). \\
    (c) The fraction of mass carried by the $b-\tau$ jet, Eq.~\ref{eq:mass_frac} ($\Delta X_{b_j \tau_h}$). In defining this observable one has used the excess of softer contamination to subjets in backgrounds as compared to the signal.\\    
    (d) N-subjettiness variable such as, $\tau_2/\tau_1$  and  $\tau_3/\tau_2$ etc. (cf. Eq.~\ref{eq:n_subjetti}) \\
     \end{enumerate}  

\begin{figure}[htb]
	\centering
		\includegraphics[scale=0.27]{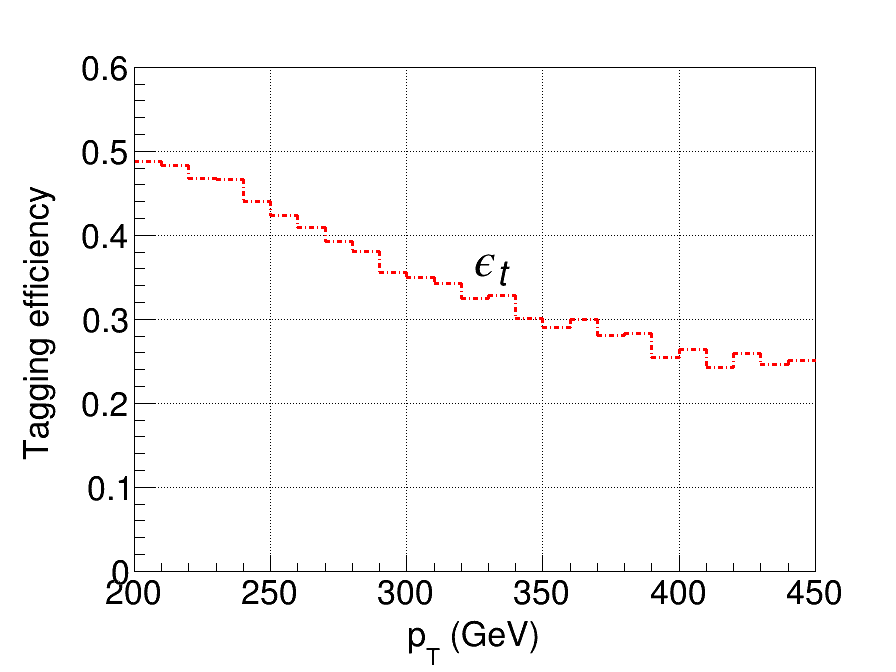}
		\caption{The efficiencies for identification of the top jet ($\epsilon_{t}$) in the $\mathrm{p_{T}}$ range of 200-450  GeV.}
        \label{fig:top_eff}
\end{figure}

\noindent Fig.~\ref{fig:top_eff} shows the efficiency $\epsilon_{t}$ of topjet identification obtained after following the above procedure (steps 1-4) for a moderate range of boosted top jet $p_T$ of 200-450 GeV. Here $\epsilon_{t}$ is defined as the ratio of number of candidate fatjet which has $b$- and $\tau$- subjets, to the corresponding total number of candidate top fatjets within matching cone of 1.5 of $t$-quark, running on all fatjets in an event. The corresponding top like jet misidentification efficiency for QCD jets is $\sim$1-2$\%$. An important observation is that $\epsilon_t$ decreases with the increasing $p_T$ of the top jet. It is due to the fact that for large values of $p_{T}$ for the fatjet, the subjets are no longer distinguishable. The tagging efficiency also depends on the radius of the fatjet.   
We have checked that by changing $\Delta R$ from 1.5 to 1.0 and the subjet radius from $0.5$ to $0.3$, the tagging efficiency increases by 5-6\%. 

We compute the efficiency of rejecting the backgrounds with the following set of cuts as described in Eq.~\ref{cut_flow_table} on the discriminators mentioned in last step above.
\begin{equation}
\begin{split}
& 50 < m_{T} < 200,\quad 0.3 < Z_{b_j} < 0.9,\quad 0.15 < Z_{\tau_h} < 0.9, \\
& 0< \tau_{2}/\tau_1 < 0.45,\quad 0.2 < \tau_{3}/\tau_2 < 0.9, \quad \Delta X_{b_j \tau_h} < 0.3\\
\end{split}
\label{cut_flow_table}
\end{equation}
With the variables mentioned in Eq.~\ref{cut_flow_table}, the signal efficiency integrated over the $p_T$ range of $t$-fatjet turns out to be 63.5\% where the corresponding mistagging efficiency due for QCD jets is 3.1\% and for $t_h \bar{t}_h$ it is 5.9\%.



\section{MVA Analysis: Unpolarized top}
\label{sec:results}

The standard cut-based strategy often rejects a significant fraction of signal events while reducing the background events   
that mimic the signal events. This situation can be improved further by employing a Multivariate analysis (MVA) \cite{Hocker:2007ht} technique which increases the background rejection rates.

We use the Boosted Decision Tree (BDT) method for the optimization purpose with the TMVA framework \cite{Hocker:2007ht}. A decision tree takes a set of input features and splits input data recursively based on those features to classify events as either signal-like or background-like. To increase the stability in the training sample with respect to statistical fluctuations, we adopt the \textit{Adaboost} \cite{FREUND1997119} technique. The decision trees are constructed using half of the signal and background events, while the other half is utilized to test the performance of the trained model. We choose the relevant hyperparameters of this method as follows: Number of trees  {\tt NTree} = 850, maximum depth of the decision tree {\tt MaxDepth} = 5, and minimum percentage of training events in each leaf node is given by {\tt MinNodeSize} = 2.5\%; other parameters are set to its default values \cite{FREUND1997119}.

Events are selected when the candidate fatjet includes a $b$- and a 
$\tau$-identified subjet inside it. 
A number of kinematic variables are constructed out of the momenta of 
these objects, as discussed in the previous section, and eventually 
10 input variables are used for BDT training. In Table \ref{tab::1}, 
the set of input variables are shown, ranking them according to 
the importance in the BDT analysis for $m_{W^\prime} = $ 1 TeV at $\sqrt{s} =13$ TeV. 
The importance here means the effectiveness of those variables in 
suppressing backgrounds while maintaining better signal purity. 
      
In a typical BDT analysis,  a few things need to be taken care of, such as instability, bias and overtraining. To that end first and foremost one has to ensure that a sufficient number of events for both signal and backgrounds are generated  such that the importance or ranking of the variables is stabilized. To remove any bias coming from a particular QCD process because of the kinematical features of the background, we optimize the number of generated events. For instance, QCD events for low $\mathrm {p_T}$ bins, such as 
 200-300 GeV regime, 10M events are simulated, 
while for 300-600 GeV and for $>$ 600 GeV regime, 
approximately 7M and 3M events are generated, respectively. We have simulated 1.3M signal events for $W^\prime$ mass = 1 TeV. In order to verify that there is no overtraining of the trees, the sample size of both training and testing data sets are optimised so that no deviation is observed in the final outcome. The goodness of fit is also checked with the Kolmogorov-Smirnov (KS) test, and observed that the KS value is within the permissible range of [0,1] and closer to the mean 0.5.    


 \begin{flushleft}
\begin{table}[ht]
 	 \centering 
 	 \begin{tabular}{l| l| l} 
 		\hline
 		Rank & Variable & Description \\ [0.5ex] 
 		\hline\hline
 		1 &  $m_T$ & Transverse mass of $b-\tau$ jets and MET of the system \\ 
 		2 & $M_{b_j\tau_h}$ & Fatjet Mass distribution in the mass range [60,160] GeV \\
 		3 & $\Delta X_{b_j\tau_h}$ & 1 - Invariant mass of $b-\tau$ jet/Fatjet mass   \\
 		4 & $\tau_2/\tau_1$ & Ratio of subjetiness of the Soft dropped Top-jet $\tau_{2}/\tau_{1}$ \\
 		5 & $Z_{b_j}$ & Fraction of energy carried by the identified $b$-jet of Soft-Dropped Top Fatjet\\
 		6 &  $Z_{\tau_h}$ & Fraction of energy carried by the identified $\tau$-jet of Soft-Dropped Top Fatjet\\
 		7 & $\tau_3/\tau_2$ &  Ratio of subjetiness of the Soft dropped Top-jet $\tau_{3}/\tau_{2}$  \\ 
 		8 & $m_{b_j}$ &  Mass of identified $b$- jet \\  
 		9 & $m_{\tau_j}$ & Mass of identified $\tau$- jet \\
 		10 & $\mathrm{N_{trk}}(\tau_h)$ & Charged track multiplicity of identified $\tau$- jet \\ [1ex]
 		\hline	
 	\end{tabular}
 \caption{List of the variables used to train $W^\prime$ signal, QCD and $t_h\bar{t}_h$ events.
 \label{tab::1}}
 \end{table}
\end{flushleft}

\begin{figure}[htb]
	\centering
		\includegraphics[scale=0.27]{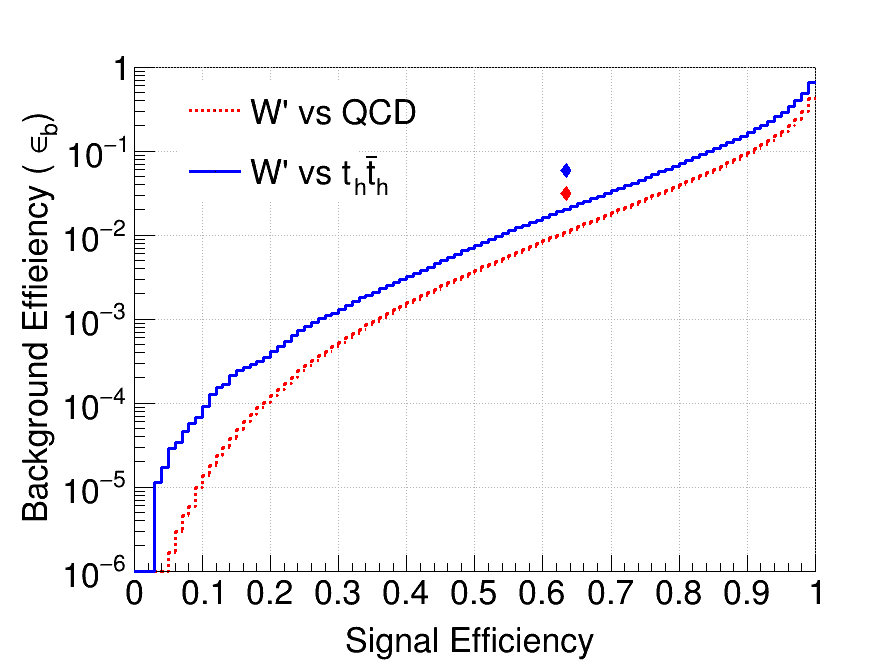}
		\caption{The Signal and background efficiencies for $m_{W^\prime} = 1 {\rm ~ TeV}$ against QCD (red dotted) and $t\bar{t}$ (blue solid). The two points in red and blue color represent the corresponding efficiencies for QCD and $t_h\bar{t}_h$ from a cut based analysis with the choice of cuts mentioned in Eq.~\ref{cut_flow_table}.}
        \label{fig:ROC}
\end{figure}

After the classifier has been trained, it gives the output in terms of a 
single variable, the BDT response. Applying a cut on the BDT output variable, 
the signal-to-background ratio is optimised and can be presented
as the Receiver Operative Characteristic (ROC). 
In Fig. \ref{fig:ROC}, we display the ROC for our proposed tagging technique. The figure estimates the tagger performance at different background rejection rates ($1 -\epsilon_b$), where $\epsilon_b$ is the background acceptance efficiency. For this classifier the 
signal-to-noise ratio is maximized at a cut value of $>$ - 0.05 where 
the signal efficiency is $\sim$ 77\% against a QCD jet mistag rate 
of $\sim$ 3\%. The figure also shows it is possible to get a good separation against the background due to hadronic top jets, with the same discriminators.

Note that, a method to identify boosted top jets consisting of electrons 
was studied in \cite{Chatterjee_2020}. 
The final outcome of our study is found to be comparable with the 
same as obtained in ~\cite{Chatterjee_2020}. We thus see that our proposed top tagger has acceptable efficiency and can be used to study boosted objects consisting of $b$ quark and $\tau$ lepton in the decays involving BSM particles in the context of BSM searches.  

\section{Tagging a polarized top}
\label{sec:polarization}

In the SM the $t/\bar t$ quarks produced via QCD are essentially unpolarized due to the vector nature of QCD whereas for the single $t$ production in association with a $W$, the $V$-$A$ nature of the $tbW$ coupling completely determines the polarization of the produced $t$. Top quarks produced from BSM sources, either in pair or singly, may have a polarization different from the predictions of the SM depending on the chiral structure of the BSM vertices responsible for its production. Hence the polarization of the produced $t$ is a good probe of many a BSM physics scenarios (see for example~\cite{BhupalDev:2007ftb,Gajdosik:2004ed,Rindani:2011pk,Belanger:2012tm,Rindani:2013mqa,Boudjema:2015nda,Godbole:2011vw,Godbole:2015bda,Arunprasath:2016tfq,Bhattacharya:2020aid,Godbole:2019erb,Belanger:2018bkh}).
As already pointed out, in the rest frame of the top quark, the angular distribution of the decay products carries information about the initial top spin direction \cite{Jezabek:1988ja,Jezabek:1994qs,Bernreuther:2008ju}. The angular distribution of the decay product (f) in the rest frame of the top with polarization $\mathcal{P}_t$, ($-1 \leq \mathcal{P}_t \leq 1$)is given by,
\begin{equation}
    \frac{1}{\Gamma}\frac{d\Gamma}{d\cos\theta_f} = \frac{1}{2}\left( 1 + k_f \mathcal{P}_t \cos\theta_f\right) ,
\label{eq:angpol}
\end{equation}
where $k_f$ is the spin analyzing power and $\theta_f$ is the angle of the decay product f with respect to the top spin direction in the top rest frame. The down-type quark and charged lepton, originating from the decay of W-boson in hadronic and leptonic top decay, respectively, have the maximum power ($k_{\ell^+} = k_{\bar{d}} = 1$). This makes the leptonic decay mode of the $t$-quark particularly suited for $t$-polarization measurements. The decay product with the next highest analysing power is the $b$-quark and one has $\kappa_b =  -\kappa_{W} = -0.41$. Note the opposite signs of $\kappa_\ell^+$ and $\kappa_b$.

In the boosted regime, the finite angular resolution of the detector reduces the effectiveness of the angular distributions of the decay products. Luckily one can construct polarization sensitive observables like energy fractions unambiguously without the requirement of W reconstruction inside the top jet \cite{Shelton:2008nq,Krohn:2009wm} to study the polarization of top quark. To this end one can exploit the kinematic features of the top decay products \cite{Shelton:2008nq,Krohn:2009wm,Kitadono:2015nxf,Godbole:2019erb,Bhattacharya:2020aid,Chatterjee_2020}. Here in this section, we explore the feasibility of distinguishing between the left-, right- and un- polarized, boosted semi-leptonic top quarks using polarization sensitive observables. For that, we repeat the procedure of top tagging, i.e steps 1-5 mentioned in the later part of Section \ref{sec:variables} and construct a few polarization sensitive variables constructed out of the energies and momenta of the tagged top jet and its subjects.

\begin{figure}[htb!]
	\begin{minipage}{0.48\textwidth}
		\centering
		\includegraphics[scale=0.27]{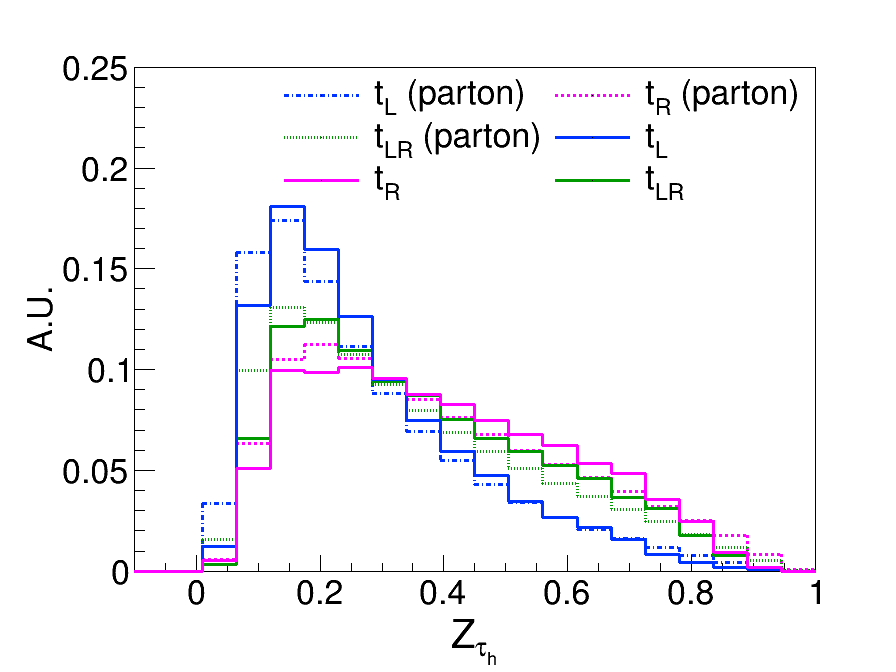}
	\end{minipage}\hfill
	\begin{minipage}{0.48\textwidth}
		\centering  
		\includegraphics[scale=0.27]{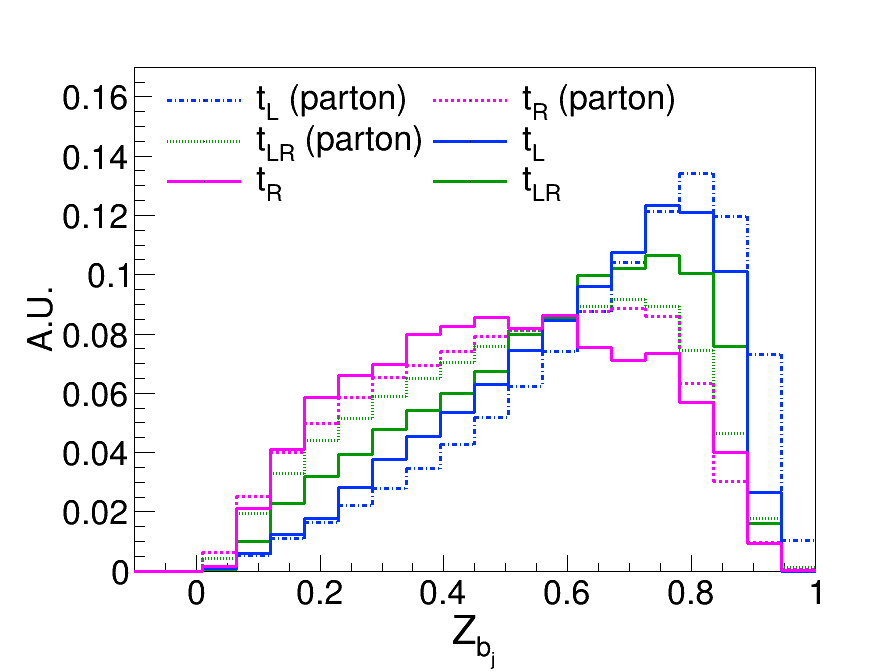}
	\end{minipage}
	\caption{Distribution of energy fraction (Eq.~\ref{eq:efrac}) of $\tau_h$- (left panel) and $b$- (right panel) like subjets for left-, right- and un- polarized top. Blue solid (dotted) lines denote the distribution for left-handed reconstructed (parton level) top and magenta solid (dotted) is for right-handed reconstructed (parton level) top. Similarly, the green solid (dotted) distributions are for reconstructed (parton level) unpolarized top.}
    \label{fig:pol_efrac}
\end{figure}

\vspace{0.3cm}
\noindent $\bullet$ {\underline{Energy Fractions}}:
\vspace{0.3cm}

The different angular distributions of the decay products in the rest frame of the $t$ get translated into different energy distributions of the decay products in the lab-frame for the boosted $t$-quark and hence of course to different distributions in energy fractions of the decay $t$ that these carry. The difference in the distributions in these energy fractions (at the parton level this is just the ratio of the lepton or $b$ quark energies to the energy of the $t$) for the left and right polarized top quarks was first pointed out in \cite{Shelton:2008nq} and then used for tagging the polarized top jet in \cite{Krohn:2009wm}. The left and right panels of Fig. \ref{fig:pol_efrac} show the behavior of energy fraction variables ($Z_{j}$) defined in Eq.~\ref{eq:efrac} for both the $\tau_h$- and $b$- like jets respectively, for the case of left-handed top quarks ${\mathrm (t_{L})}$, right-handed top quarks ${\mathrm (t_{R})}$ and unpolarized top quarks ${\mathrm (t_{LR})}$ originating from $W^\prime$ decay. The corresponding expected distributions at the parton level are shown by dotted lines for comparison. For a more realistic comparison with the jet level plots, one uses the visible energy from $\tau$-lepton to determine $Z_{\tau_h}$ as described in Sec.~\ref{sec:variables}.

We do see the same difference, as seen in Fig.~\ref{fig:efrac}, between the energy fractions carried by the $b$-jets and $\tau_h$-jets. Further, we note that $b$-like jets from $t_L$ are more boosted compared to those from the $t_R$ whereas for the $\tau_h$, exactly opposite is the case. This can be of course understood in terms of the opposite signs of the spin analysing powers of the $b$ quark and the $\tau^+$. Due to the negative sign of $\kappa_b$ according to Eq. \ref{eq:angpol}, $b$ quarks are preferentially emitted opposite to the spin direction in the rest frame of the $t$. For the $t_{L}$, this means that they are emitted preferentially in the direction of motion of the $t$ in the laboratory. This, in turn, means that the boost from the rest frame of the $t$ to the laboratory frame makes these $b$ quarks more energetic than would be the case with $b$ quarks coming from the decay of $t_{R}$ or unpolarised $t$-quarks. For the $\tau^+$ the positive nature of $\kappa_\tau$ implies exactly the opposite. As can be seen from the left panel of Fig.~\ref{fig:pol_efrac}, the $\tau$-jet energy fraction ($Z_{\tau_h}$) peaks $\sim 0.15$ for $t_{L}$, taking relatively larger values ($\sim 0.3$ or more) for $t_{R}$. The right plot of Fig.~\ref{fig:pol_efrac}, displays the characteristics of the $b$-jet that it takes away more energy for $t_{L}$ than $t_{R}$. 
The distributions $Z_{b_j}$ shown in the right panel of Fig.~\ref{fig:pol_efrac} peak at $\sim 0.8 ~\&~ 0.4$ for $t_{L}$ and $t_{R}$ respectively. For comparison, we have included the distributions of energy fractions ($Z_{\tau_h}$ and $Z_{b_j}$) for $t_{LR}$ which we have previously discussed in Fig.~\ref{fig:efrac}. Thus, we see that even though the full reconstruction of the $\tau$ momentum is not possible at the jet level (due to the presence of neutrinos in the final state), the energy fractions can act as a good polarimeter for differentiating left-handed, right-handed and unpolarized boosted top quarks decaying to final state containing $\tau$ leptons.

\vskip 0.3cm
\noindent
$\bullet$ \underline{Angular variable (${\rm {cos~\theta_j}}$)}:
\vskip 0.3cm
Eq. \ref{eq:angpol} gives the distribution in the angle of the decay product with the spin direction of the decaying $t$, in the rest frame of the $t$. Since for the $t$-quark of a given helicity, the spin direction is related to the direction of the $t$ three-momentum vector; we can instead look at the correlation of the decay product momentum in the rest frame of the $t$, with the original $t$ direction in the laboratory. In case of the hadronic topjet, the topjet direction in the laboratory is of course a very good proxy for the momentum of the $t$-quark in laboratory. This was used to good effect in \cite{Godbole:2019erb} to discriminate boosted topjets with different helicities. In the present case due to the presence of missing momentum, we consider sum of the reconstructed momentum of the $b-\tau$ jet system in the laboratory frame, viz. $(\vec{b_j} + \vec{\tau_h})$ system as a proxy for the $t$ momentum (and hence top-spin) direction. We then define
\begin{equation}
{\rm cos \theta_J} \big\vert_{J = b_j,\tau_h} = \frac{(\vec{b}_j + \vec{\tau}_j). \vec{j}^\prime}{\left|(\vec{b}_j + \vec{\tau}_j)\right| \left|\vec{j}^\prime\right|}.
\label{eq:costheta}
\end{equation}
Here $j^{\prime}$ is the momentum of the subjet (either $b$-jet or $\tau$-jet) in the rest frame of the $b$-$\tau$ jet system. 

\begin{figure}[htb!]
	\begin{minipage}{0.48\textwidth}
		\centering
		\includegraphics[scale=0.27]{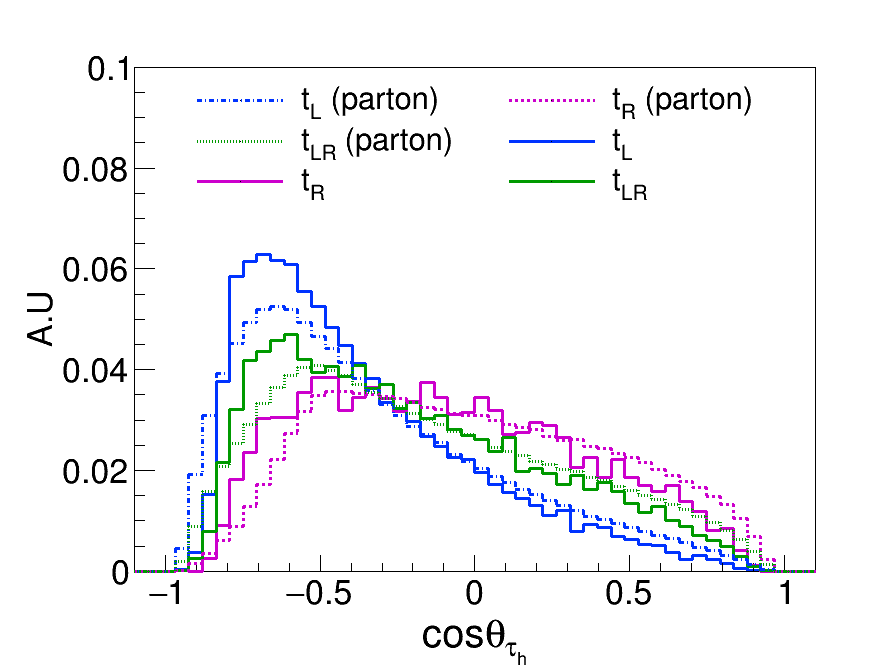}
	\end{minipage}\hfill
	\begin{minipage}{0.48\textwidth}
		\centering  
		\includegraphics[scale=0.27]{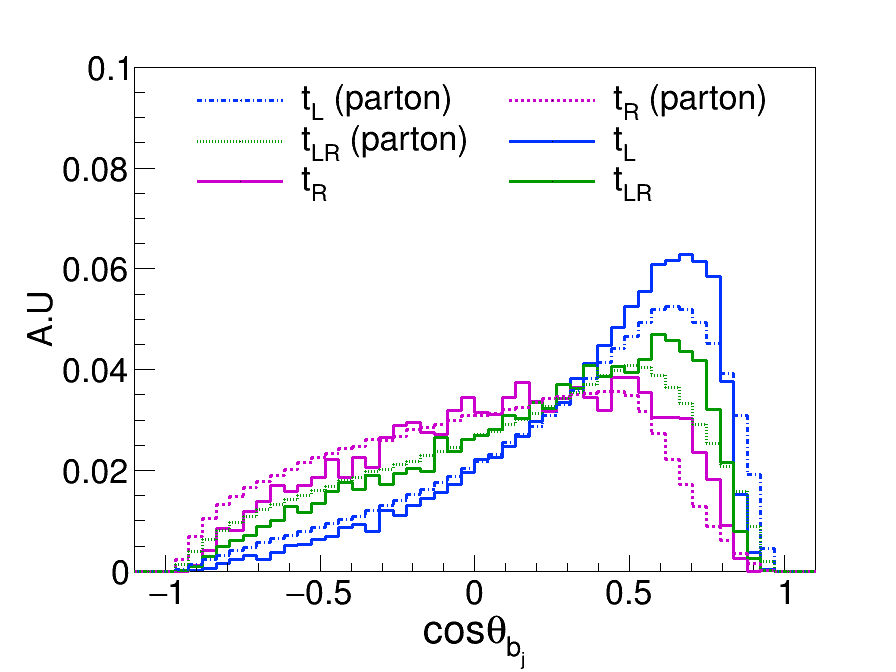}
	\end{minipage}
	\caption{The angular distribution of $\tau_h$ (left) and $b_j$ (right) in the rest frame of $(b_j+\tau_h)$ system for $t_L$, $t_R$ and $t_{LR}$ following Eq. \ref{eq:costheta}. The color code is similar to Fig.~\ref{fig:pol_efrac}.}
	\label{pol:costhetastar}
\end{figure}

For left-handed top quark, the direction of the top quark spin is in the opposite direction with respect to the top quark momentum in the laboratory frame. Eq. \ref{eq:angpol} and the values of $\kappa_{\ell^+}, \kappa_b$, tell us then that the $\tau_h (b)$ from $t_R$ would be preferentially emitted along (opposite) the direction of the $(\vec{b_j} + \vec{\tau_h})$, i.e. the proxy $t$ momentum direction. Exactly opposite will be the case for $t_L$, and for $t_{LR}$ it will be in between $t_L$ and $t_R$. This is borne out by the left plot in Fig.~\ref{pol:costhetastar}, which shows that the $\cos \theta_{\tau_h}$ is preferentially positive (negative) for right (left) handed top quarks. As seen in the right plot in Fig.~\ref{pol:costhetastar}, the $b$-jet exhibits a behavior exactly opposite to the $\tau$-case i.e.,
$\cos \theta_{b_j}$ takes preferentially negative (positive) values for right (left) handed top quark.         

\begin{figure}[htb!]
	\centering
		\includegraphics[scale=0.27]{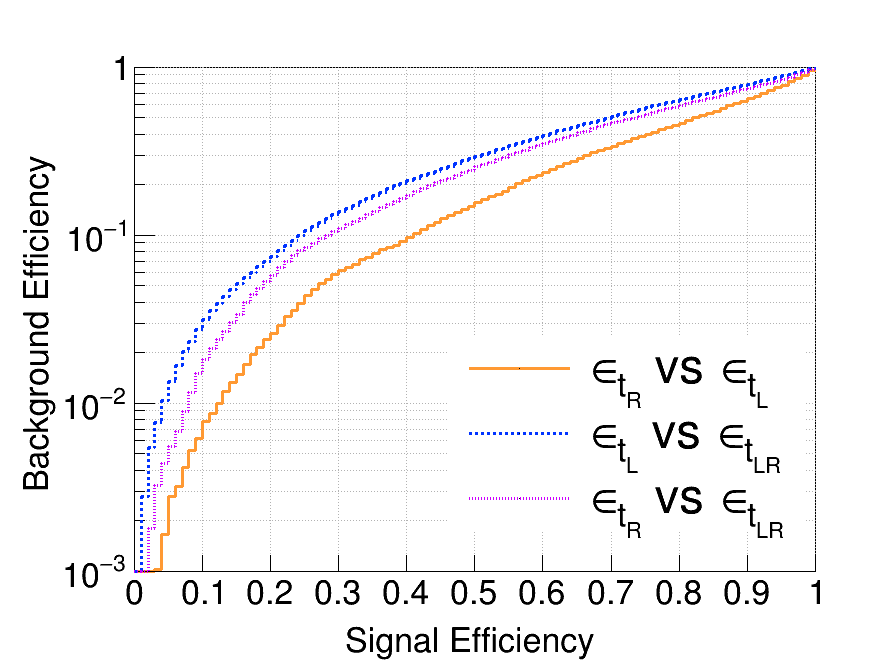}
		\caption{The ROC curve estimates the performance of the BDT classifier of distinguishing the (a) right vs left (b) left vs unpolarized (c) right vs unpolarized top jets.}
		\label{fig:pol_ROC}
\end{figure}

Using only these two sets of variables, namely the energy fractions ($Z_{b_j}$ and $Z_{\tau_h}$) and angular correlation discriminator ($\cos\theta_j$) of the top decay products, we can achieve a good separation between the left-handed, right-handed and unpolarized top quarks. Note that the top tagging efficiencies for left and right polarized top quark do differ only slightly $\sim 2-3 \%$. In Fig.~\ref{fig:pol_ROC}, we display the ROC curve obtained by training the BDT with two polarization sensitive observables along with the other variables of Table \ref{tab::1}, for left, right and unpolarized top samples. The blue (magenta) curve shows the efficiency of $t_L$ ($t_{R}$) against $t_{LR}$, while the yellow line represents the same for $t_R$ against $t_L$. 
It can be seen from Fig.~\ref{fig:pol_ROC} that $t_R$ is identified more efficiently than $t_L$ against an unpolarized top background. 
This can be understood by looking at kinematic regions where one of the $t_L$ or $t_R$ differs distinctly from $t_{LR}$. In this case, the $b$-jet energy fraction seems to be a better discriminator between the polarized top and unpolarized sample for the right polarized top. As we pointed out previously, because a fraction of the energy of top is carried out by the MET in $\tau$ decay, $Z_{b_j}$ peaks at higher values than $Z_{\tau_h}$ for $t_{LR}$. Hence, $t_{R}$ for which the b quark from t decay is less energetic than that of $t_{L}$, has a higher chance of discrimination against $t_{LR}$.
We find that right-handed top jets can be tagged with $\sim$ 65\% efficiency ($\epsilon_{t_R}$) with a mis tagging rate of $\sim 25\%$ for left-handed top jets ($\epsilon_{t_L}$) and $\sim 35\%$ against unpolarized top ($\epsilon_{t_{LR}}$) jets. If we reduce $\epsilon_{t_L}$ to $\sim 1-2\%$, $\epsilon_{t_R}$ comes out to be around $15-20\%$. 

\section{Summary}
\label{sec:summary}
In this paper, we investigated the performance of a boosted top tagger when the top quark decays in the leptonic channel with the $\tau$ lepton in the final state. 
The proposed top tagger relies on the identification of an energetic $b$-jet and a $\tau$-jet within the top fatjet and its energy profile with respect to the parent particle. This methodology focuses on the distribution of energy between two subjets within the fatjet and constructs kinematic variables 
relevant to the final state topology. This in turn helps to discriminate the signal from SM backgrounds. Some of the variables that are found to be useful in identifying and classifying the semi-leptonic boosted tops are (a) the energy fraction of the identified $b$- and $\tau$- subjets of the total fatjet energy, (b) the difference of the masses of the $b$-$\tau$ system and top fatjet (cf. Eq.~\ref{eq:mass_frac}), (c) the ratios of the N-subjettiness variables (d) transverse mass of the $b$-$\tau$ subject.

We analyze the performance of the proposed tagger by using simulated signal and background events and then constructing observables based on the jet substructure technique. Through a BDT analysis, we obtain a signal efficiency of around 77\% while
keeping the mistagging rate of the QCD jets (consists of quark and gluon initiated jets) to 3\% level. The hadronically decaying top quark initiated jets can also play the role of a potential background. We find that with a signal efficiency of around 77\%, the mistagging rate of the hadronic top quark jet is around 5-6\%. It is noteworthy that even though the main focus of this study is to develop a toptagger when top quark decays through $\tau$-lepton in the final state,  however, this tagger can be applied to any fatjet, which includes a $b$-jet and a $\tau$-jet. For example, decays of a $3^{\text{rd}}$ generation Leptoquark, top squark decay in R-parity violating supersymmetric models etc. can all lead to a final containing a $b$- and a $\tau$- subjet inside a fatjet due to the boosted nature of the parent particle. Absence of a $\nu$ in the primary decay in the signal ought to make the tagging using this strategy even easier.


Another important aspect of developing a toptagger for this leptonic final state, is that we can analyze and estimate the sensitivity of the tagger with respect to top quark polarization. The couplings of the top quark with fermions and bosons in various BSM scenarios can have different implications for the top quark polarization. This polarization in turn modifies the distribution of various kinematic observables involving its decay products. We study the sensitivity of these distributions to $t$-polarization in two extreme cases of top polarization, namely purely left-handed and purely right-handed top quarks. We use the same method of top-tagging as described above. We indeed find that using the observables based on the energy profile of subjets of the boosted top jet, namely the $b$- and $\tau$- tagged jets along with the angular correlations measured in the rest frame of $b$-$\tau$ system, one can differentiate between the left and right-handed top quarks quite efficiently. A detailed exploration of some of the interesting applications of our proposed top tagger is left for future investigations.

\section*{Acknowledgements}
The work of AC is funded by the Department of Science and Technology, Government of India, under Grant No. IFA18-PH 224 (INSPIRE Faculty Award). RMG wishes to acknowledge the support of Indian National Science Academy under the award of INSA Senior Scientist. A.D. thanks Ritesh Kumar Singh (IISER, Kolkata) for helpful discussions. A.D. would like to acknowledge the support of INSPIRE Fellowship IF160414.

\bibliography{References}
\bibliographystyle{unsrturl}


\end{document}